\documentclass[final,5p,times,twocolumn]{elsarticle}
 \biboptions{comma,sort&compress}
\usepackage{graphicx}
\usepackage{here}

\def\beq{\begin{equation}}
\def\eeq{\end{equation}}
\def\bea{\begin{eqnarray}}
\def\eea{\end{eqnarray}}
\def\bq{\begin{quote}}
\def\eq{\end{quote}}

\def\nnb{\nonumber}
\def\ga{\left(}
\def\dr{\right)}

\def\lb{\lbrack}

\def\rar{\rightarrow}

\def\nnb{\nonumber}
\def\la{\langle}
\def\ra{\rangle}
\def\nin{\noindent}
\def\ba{\vspace*{-0.2cm}\begin{array}}
\def\ea{\end{array}\vspace*{-0.2cm}}

\def\b{$\bullet~$}
\def\als{\alpha_s}

\def\gg2{ \la\alpha_s G^2 \ra}
\def\gg3{g^3f_{abc}\la G^aG^bG^c \ra}
\def\ggg4{\la\als^2G^4\ra}

\def\lb{\label}

\journal{Physics Letters B}

\begin{document}

\begin{frontmatter}

\title{
Gluon condensates
and  $\overline{m}_b(\overline{m}_b)$ from QCD-exponential  sum rules at higher orders
} 

 \author[label1]{Stephan Narison
 }
   \address[label1]{Laboratoire
Univers et Particules , CNRS-IN2P3,  
Case 070, Place Eug\`ene
Bataillon, 34095 - Montpellier Cedex 05, France.}
\ead{snarison@yahoo.fr}


\pagestyle{myheadings}
\markright{ }
\begin{abstract}
\noindent
We test the convergence of the QCD exponential sum rules by
including PT corrections to order $\alpha_s^3$ and the NP contributions up to dimension $D=8$ condensates. Then, using the ratio of exponential sum rules where the QCD PT series is more convergent, we study the correlation  between the gluon condensates $\la \alpha_s G^2\ra$ and $\la g^3f_{abc} G^3\ra$. From charmonium systems and using the charm quark mass as input, we deduce:
$\la g^3f_{abc} G^3\ra=(8.2\pm1.0)~{\rm GeV}^2\times \la \alpha_s G^2\ra$ corresponding to  $\la \alpha_s G^2\ra=(7.5\pm 2.0)\times 10^{-2}$ GeV$^4$. Using these results for the
 bottomium systems, we obtain: $\overline{m}_b(\overline{m}_b)=  4212(32)$ MeV, which is slightly higher but consistent within the errrors with the ones from $Q^2$-moments and their ratios $\overline{m}_b(\overline{m}_b)=  4172(12)$ MeV. 
 We are tempted to consider as a final result from the sum rules approaches, the average $\overline{m}_b(\overline{m}_b)=  4177(11)$ MeV 
 of the two previous determinations.
\end{abstract}

\end{frontmatter}
\section{Introduction and motivations}
\vspace*{-0.25cm}
 \nin
\b In recent letters \cite{SNcb,SNcb2}, we have used  finite $n$ and $Q^2$-moments ${\cal M}_n(Q^2)$:
 \bea
 {\cal M}_n\ga -q^2\equiv Q^2\dr&\equiv& 4\pi^2{(-1)^n\over n!}\ga {d\over dQ^2}\dr^n \Pi(-Q^2)\nnb\\
 &=&\int_{4m_Q^2}^\infty dt {{R}(t,m_c^2)\over (t+Q^2)^{n+1}}~,
 \eea
 for
extracting the values of the gluon condensates and precise values of the running $c,b$ quark  masses in the
$\overline{MS}$-scheme. The results for the condensates are:
\bea
\la \alpha_s G^2\ra&=&(0.07\pm 0.01)~{\rm GeV}^4~,\nnb\\
\la g^3f_{abc} G^3\ra&=&(8.8\pm 5.5)~{\rm GeV}^2\times \la \alpha_s G^2\ra~,
\lb{eq:cond_mom}
\eea
and for the quark masses (average of the two consistent determinations in \cite{SNcb} and \cite{SNcb2}):
\bea
\overline{m}_c(\overline{m}_c)&=&  1262(17)~{\rm MeV}~,\nnb\\
\overline{m}_b(\overline{m}_b)&=&  4172(12)~{\rm MeV}~.
\lb{eq:mass_mom}
\eea
\b Following the same lines, we study in this letter the QCD-exponential moments:  
 \beq
 {\cal L}_p\ga \tau\dr=\int_{4m_Q^2}^\infty dt~ t^p~e^{-t\tau}{R_{e^+e^-}(t,m_c^2)}~~~{\rm for}~~~p=0~,
 \lb{eq:mom}
 \eeq
 and its ratio:
 \beq
 {\cal R}_0(\tau)\equiv-{d\over d\tau} {\rm Log} {\large{[}{\cal L}_0\large{]}} =  { {\cal L}_{1}\over  {\cal L}_0}~,
 \lb{eq:ratio}
\eeq
 when higher order PT and NP terms are included into their QCD expressions; 
 $R_{e^+e^-}$ is  the ratio  of the total cross-section of $\sigma(e^+e^-\to$ hadrons) over $\sigma(e^+e^-\to \mu^+\mu^-)$, which is normalized asymptotically to one;  $\tau$ is the sum rule variable. \\
 \b However, working with higher values of $p$ does not help due to the relative increase of the higher energy contributions to the spectral integral which are not under a good control and which decreases the accuracy of the predictions. Then, we shall limit ourselves to the case $p=0,1$ where  ${\cal L}_0$ can be derived naturally from the moments ${\cal M}_n\ga  Q^2\dr$ by taking the limit: 
  \beq
 n\rar\infty~~{\rm and}~~ Q^2\rar \infty ~~~{\rm but~~ keeping}~~ {n\over Q^2}\equiv \tau ~~{\rm finite}~.
 \label{eq:limit}
 \eeq
\b Therefore,  by working with $ {\cal L}_0$ and $ {\cal L}_1$, one can explore a new region of energy which is not reached 
 with ${\cal M}_n\ga  Q^2\dr$ at finite  and relatively small $n$ and $Q^2$ values like e.g. the ${\cal M}_n\ga  Q^2=0\dr$ moments. One can also notice that in the case of the exponential sum rules, $Q^2$ and $n$ are correlated via the sum rule variable $\tau$, which is not the case of  ${\cal M}_n\ga  Q^2\dr$ where they are completely free and uncorrelated. 
 In this sense, the two approaches are different but complementary and
 one needs to study both methods for a complete test of the sum rule approaches. \\
\b Originally named Borel sum rules by SVZ because of the appearance of  a factorial suppression factor in the non-perturbative condensate contributions, it has been shown by \cite{SNR} that the PT radiative corrections satisfy instead the properties of an inverse Laplace sum rule. Exponential sum rules have been used successfully by SVZ for light quark systems \cite{SVZ}. Especially, the ratio of moments has been largely used in the literature as it appears to be a useful tool for extracting the masses of lowest hadrons \cite{SNB1,SNB2,SNB3}. \\
\b In the case of heavy quarks,  Bell-Bertlmann have used the exponential sum rules (so-called magic moments) in a series of papers  for their relativistic and non-relativistic versions \cite{BELL,BERT,NEUF}.  In particular, they have studied its quantum mechanical interpretation using the model of harmonic oscillator where they found that the sum rule variable $\tau$ is related to the imaginary time while the optimal value of the results can be extracted at the minimum or inflexion point (stability point) of the theoretical curves. Using this method, it has been emphasized by \cite{BELL,BERT,NEUF} that the value of the gluon condensate $\la \alpha_s G^2\ra$ obtained by SVZ \cite{SVZ} has been underestimated by about a factor 2. A result which has been supported by forthcoming papers in different channels \cite{PEROTTET,SNI,SNGh,SHAW}. \\
\b Unfortunately, the existing different applications of the exponential Borel/Laplace sum rule for heavy quarkonia systems \cite{BELL,BERT,NEUF,SHAW} and for heavy exotic states (see e.g. \cite{EXOTIC}) suffer from the ambiguous definition of the heavy quark masses used in the analysis because, in these applications, the QCD series are only known to lowest order or at most to order $\alpha_s$. This feature does not permit to extract accurate predictions of the quark and hadron masses from exponential sum rules, with the exception of working with double ratio of sum rules (DRSR) introduced in \cite{DRSR} for extracting mass-splittings and ratio of couplings and of form factors despite the poor knowledge of the QCD corrections in these channels because the systematics of the method and QCD corrections tend to disappear \cite{SNGh,SNFBS,SNhl,SNme+e-,mnnr,HBARYON,drx,SNFORM}.  \\
\b In this letter, we shall improve the previous old extractions done 20-30 years ago of the PT (heavy quark masses) and NPT (gluon condensates)  QCD parameters  \cite{BELL,BERT,NEUF,SHAW,SNmc} by including into the
exponential sum rules, pQCD corrections up to order $\alpha_s^3$ and non perturbative QCD condensates up to dimension $D=8$ condensates. Unlike Refs. \cite{BELL,BERT,NEUF,SHAW}, we shall not also use the input value of the $\la g^3f_{abc} G^3\ra$ deduced from the Dilute gas Instanton estimate but leaves it as a free parameter in the analysis. In particular, it is important to check these new higher order effects on the estimate of the gluon condensate  $\la \alpha_s G^2\ra$ and of the quark masses done previously by Bell-Bertlmann \cite{BELL,BERT,NEUF}, which we shall compare with some new results obtained in \cite{SNcb,SNcb2} using finite $Q^2$ and $n$-moments ${\cal M}_n\ga  Q^2\dr$. Indeed, a reliable control of these
QCD parameters are mandatory for further applications of the SVZ sum rules. \\
\b By parametrizing the high-energy part of the spectral function from a threshold $t_c$ by a ``QCD continuum" 
which comes from the discontinuity of the Feynman diagrams contributions to the two-point function,
one can transform the previous sum rules in Eqs. (\ref{eq:mom}) and (\ref{eq:ratio}) into  Finite Energy Sum Rules (FESR):
\beq
 {\cal L}^{t_c}_0\ga \tau\dr=\int_{4m_Q^2}^{t_c} dt ~e^{-t\tau}{R_{e^+e^-}(t,m_c^2)}~, ~~~~{\cal R}_0^{t_c}(\tau)={ {\cal L}^{t_c}_{1}\over  {\cal L}^{t_c}_0}\simeq M^2_R,
 \lb{eq:fesr}
 \eeq
where $M_R$ is the lowest resonance mass. 

\section{Expressions of the sum rules }
\label{sec:sumrule}
\vspace*{-0.25cm}
 \nin
\b We parametrize the spectral function using the $J/\psi$ family masses and widths for the $c$-quark channel and the ones of the $\Upsilon$ family in the $b$-quark channel. These inputs are given in Tables 1 and 3 of \cite{SNcb}. We add to these resonance contributions the one of the QCD continuum
above respectively the threshold $\sqrt{t_c}=(4.6\pm 0.1)$ and $(11.098\pm 0.079)$ GeV, which is described by the asymptotic pQCD expression in the massless quark limit in the $\overline{MS}$-scheme,  which reads for $n_f=4$ flavours \footnote{Original papers are in  Refs. 317 to 321 of the book \cite{SNB1}.}:
\beq
R_{e^+e^-}\vert_{cont}=1+a_s+1.5a_s^2-12.07a_s^3~,
\eeq
where \footnote{In the following, we choose the subtraction point $\nu=\overline{m}_Q$, where all quantities will be evaluated.}:
\beq
a_s\equiv {\alpha_s(\overline{m}_Q)/\pi}~.
\eeq
\b The perturbative QCD expression is deduced from the well-known spectral function to order $\alpha_s$ within  the on-shell renormalization scheme \cite{KALLEN,SCHWINGER}, which we have transformed to the $\overline{MS}$-scheme using the relation between the on-shell and $\overline{MS}$-mass truncated at the same order of PT series \cite{SNMass}:
\beq
M_Q=\overline{m}_Q(\overline{m}_Q)\ga 1+{4\over 3}a_s\dr~.
\eeq
\b We add to this expression the PT approximate spectral function to order $\alpha_s^2$ in the $\overline{MS}$-scheme \cite{CHET3} and we add to it the one of $\alpha_s^3$ term deduced from the expression  given in \cite{HOANG}. \\
\b The $D=4$ gluon condensate contribution comes from SVZ \cite{SVZ} to LO and from \cite{BROAD} to NLO. \\
\b The $D=6$ and 8 contributions come from \cite{NIKOL,NIKOL2}. Their analytic contributions to the the ratio of exponentiel moments are given by \cite{NEUF}. We shall use the modified factorization of the $D=8$  proposed by \cite{BAGAN} based on $1/N$ expansion with a value of about 1/2 of the one from a na\"\i ve factorization \cite{NIKOL2,NSVZ} which is favoured by the analysis of the $Q^2$-moments in \cite{SNcb}.
\section{Tests of the convergence and choice of the sum rules }
\label{sec:converge}
\vspace*{-0.25cm}
 \nin
 In the following, we test the convergence of the charmonium sum rules which will permit us to select the quantity
which can provide reliable results.  \\
\begin{figure}[hbt]
\begin{center}
\includegraphics[width=7cm]{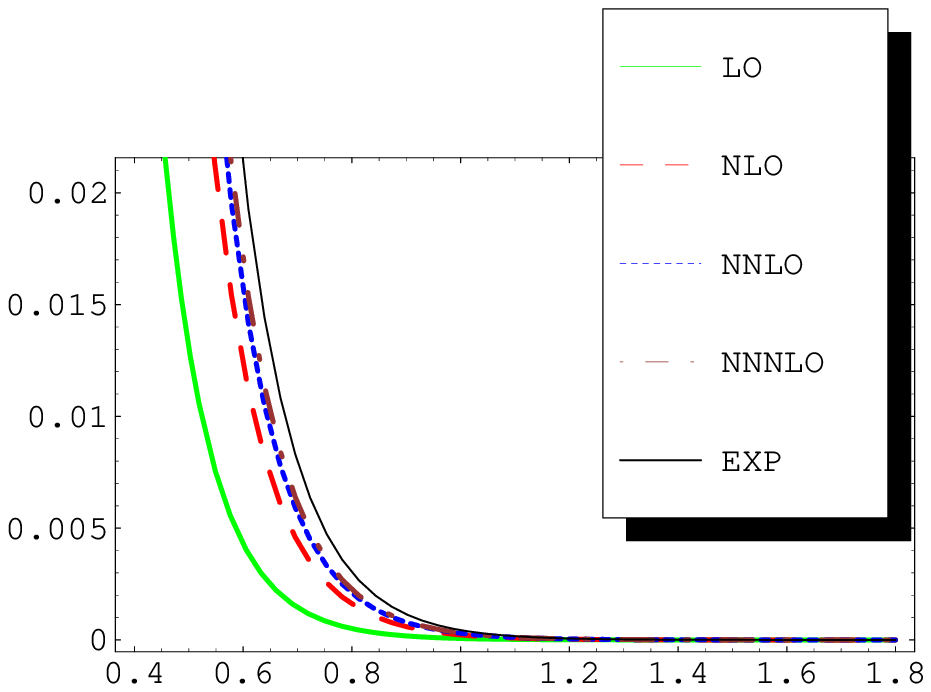}
\includegraphics[width=7cm]{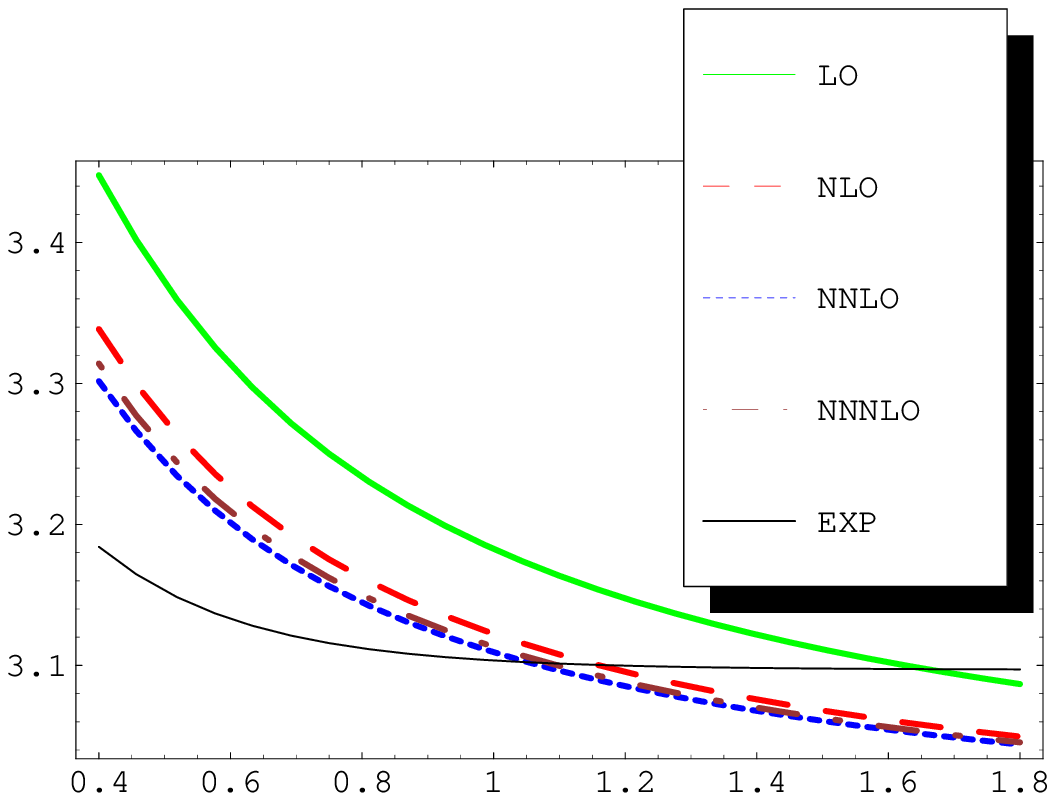}
\caption{\footnotesize {\bf a)} Behaviour of the moment ${{\cal L}^c_0}(\overline{m}_c^2)$ in GeV$^2$ versus $\tau$ in GeV$^{-2}$
at different orders of perturbation theory: LO: green (thick continuous); NLO: red (long dashed); NNLO: blue (short dashed); NNNLO: red-wine (dot-dashed).  The  black (continuous) curve comes from the central values of the data including the QCD continuum. {\bf b)} The same as in a) but for the ratio of moments $\sqrt{{\cal R}^c_0}(\overline{m}_c^2)$.
} 
\label{fig:pert}
\end{center}
\end{figure} 
\\
 \nin
\b Given the value of $\overline{m}_c(\overline{m}_c)$ in Eq. (\ref{eq:mass_mom}), one can deduce the numerical PT series normalized to the LO contribution at the subtraction point $\nu=\overline{m}_c^2$ and taking e.g. $\tau=1.35$ GeV$^{-2}$ where the ratio of sum rules stabilizes in $\tau$ (see Fig. \ref{fig:ratioc}):
\beq
{\cal L}_0(\overline{m}_c^2)\vert_{PT}\simeq   1+22.6a_s+85.2a_s^2+152.8a_s^3~,
\eeq
where one can notice that the PT corrections are large.  For the ratio of moments normalized to the lowest order expression, we obtain within the same approximation and for $\tau=1.35$ GeV$^{-2}$:
\beq
\sqrt{{\cal R}_0}(\overline{m}_c^2)\vert_{PT}\simeq 1-0.12a_s-0.17a_s^2+0.57a_s^3~,
\eeq
where in this case, the PT corrections
are much smaller and the series converge quite well. We illustrate the previous discussions by showing the $\tau$-behaviours of the PT moments and of their ratio in Fig. \ref{fig:pert}. We also show the experimental part of the moments (continuous-black curve). We have used the central values of the input QCD parameters in Eqs. (\ref{eq:cond_mom}) and (\ref{eq:mass_mom}) and the value of $\alpha_s$ deduced from the one from $\tau$-decays \cite{SNTAU,BNP}:
\beq
\alpha_s(\overline{m}_c)=0.408(14)~.
\eeq
\b At this value of $\tau$, the non-perturbative contributions to the ratio of moments normalized to the PT LO term read:
\bea
\sqrt{{\cal R}_0}(\overline{m}_c^2)\vert_{NP}&\simeq& 1.93\la \alpha_s G^2\ra \ga 1-3.23a_s\dr \tau^2\nnb\\
&&-0.078 \la g^3 f_{abc}G^3\ra\tau^3+0.048\tau^4~,
\eea
which also indicates a quite good convergence of the OPE despite this relatively large value of $\tau$. \\ \b Therefore, in the following, we shall definitely work with the ratio of moments for extracting 
the QCD parameters. 
\section{$\la g^3 f_{abc}G^3\ra$ versus $\la \alpha_s G^2\ra$ from charmonium  }
\label{sec:g3}
\vspace*{-0.25cm}
\begin{figure}[hbt]
\begin{center}
\includegraphics[width=8.cm]{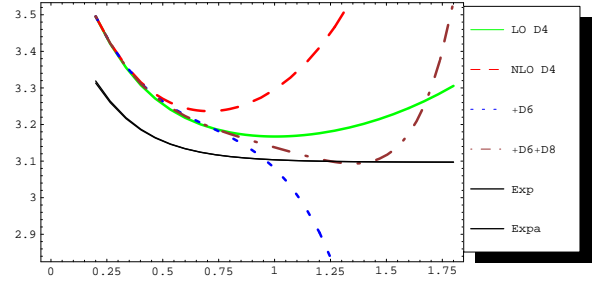}
\caption{\footnotesize Behaviour of the ratio of moments $\sqrt{{\cal R}^c_0}(\overline{m}_c^2)$ in GeV versus $\tau$ in GeV$^{-2}$. The two black curves almost degenerate are the data including error bars. The green (thick continuous) is the PT contribution including the $D=4$ condensate to LO. The long dashed (red) curve is the contribution including the $\alpha_s$ correction to the $D=4$ contribution. The short dashed (blue) curve is the QCD expression including the $D=6$ condensate and the dot-dashed (red-wine) is the QCD expression including the $D=8$ contribution. } 
\label{fig:ratioc}
\end{center}
\end{figure}
\begin{figure}[hbt]
\begin{center}
\includegraphics[width=7cm]{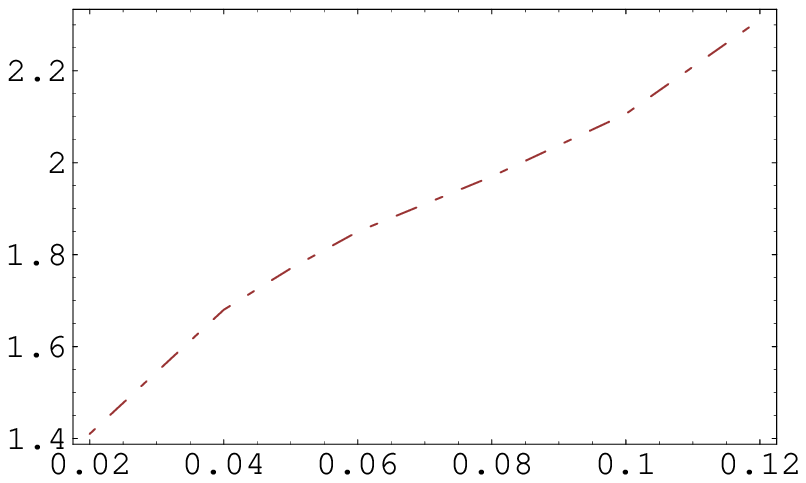}
\caption{\footnotesize Variation of $\rho_c$ in GeV$^{-1}$ versus $\la\alpha_s G^2\ra$ in GeV$^4$ from the ratio of moments ${\cal R}^c_0(\overline{m}_c^2)$} 
\label{fig:rhoc}
\end{center}
\end{figure}

 
 \nin
 In this section, our aim is to improve the estimate of the triple gluon condensate $\la g^3 f_{abc}G^3\ra$
 from the ratio of the exponential sum rules using its correlation with the gluon condensate $\la \alpha_s G^2\ra$ by giving as input the  precise value of the charm quark mass $\overline{m}_c(m_c)$ in Eq. (\ref{eq:mass_mom})
  obtained in \cite{SNcb,SNcb2} from moment sum rules. \\
\b We show in Fig. \ref{fig:ratioc}, the $\tau$-behaviour of $\sqrt{{\cal R}^c_0}(\overline{m}_c^2)$ by using the central values of the QCD parameters in Eq. (\ref{eq:mass_mom}) and the best central value of the instanton radius $\rho_c=1.89$ GeV$^{-1}$ obtained in Eq. (\ref{eq:rhoc}) by adjusting the minimum of $\sqrt{{\cal R}^c_0}(\overline{m}_c^2)$ to the experimental data. One can notice that at the minimum, one has:
\beq
\sqrt{{\cal R}^c_0}(\overline{m}_c^2) \vert_{min}=M_{J/\psi}~,
\eeq
indicating that the higher state contributions to the ratio of moments are neglible at this value of $\tau$. \\
\b  We show in Fig. \ref{fig:rhoc} the correlation between $\rho_c$ and
$ \la \alpha_s G^2\ra$ from which we can find a slight stability (inflexion point) for the range:
\beq
 \la \alpha_s G^2\ra=(7.5\pm 2.0)\times 10^{-2}~{\rm GeV}^4~,
 \eeq
 which is comparable with the one  in Eq. (\ref{eq:cond_mom}) obtained from the $Q^2$-moments though less accurate here. In this region, we can deduce in GeV$^{-1}$\,\footnote{If we have used the value of $\la\alpha_s G^2\ra$ in Eq. (\ref{eq:cond_mom}) from the moments, we would have obtained a more precise value of $\rho_c$.}:
\beq
\rho_c=1.89~(10)_{ G^2}~(1)_{\alpha_s}~(5)_\nu~(1)_\tau~=1.89\pm 0.11~,
\lb{eq:rhoc}
\eeq
where the error comes respectively from $\la\alpha_s G^2\ra$, $\alpha_s$, the choice of subtraction point $\nu$, the localisation of the minimum : $\tau\simeq (1.25\sim 1.45)$ GeV$^{-2}$. We have estimated the effect of the subtraction point $\nu$ by varying it from $\overline{m}_c(m_c)$ to $M_\tau$ and using the substitution (see e.g. \cite{SNB3,SNB1}):
\bea
\alpha_s(m_c)&\to& \alpha_s(\nu)\times \ga 1-{\beta_1}~{\alpha_s(\nu)\over\pi}\log{\nu\over  m_c}\dr~,
\label{eq:sub}
\eea
where $\beta_1=-(1/2)(11-2n_f/3)$ for $n_f$-flavours. The error due to the $\alpha_s^{n\geq 4}$ is obtained by taking the $\pm$ sign in the coefficient of $\alpha_s^3$ and is negligible. The error due to the data is also negligible.\\
\b The previous value of $\rho_c$ in Eq. (\ref{eq:rhoc}) leads to:
\beq
\la g^3f_{abc} G^3\ra=(8.2\pm 1.0)~{\rm GeV}^2\times \la \alpha_s G^2\ra~,
\lb{eq:cond}
\eeq
where the relation from the Dilute Gas Instanton approximation:
\beq
{\la g^3f_{abc}G^3\ra\over \la \alpha_s G^2\ra}= {4\over 5}{12\pi\over\rho_c^2}~,
\label{eq:conda}
\eeq
has been used.  
We consider this result as
an improvement of the one in Eq. (\ref{eq:cond_mom}) from the $Q^2$-moments \cite{SNcb}. Notice that in the past \cite{BELL,BERT,NEUF,SHAW}, the contribution of $\la g^3f_{abc} G^3\ra$ has been negligible because of the (a priori) use of the instanton radius $\rho_c\simeq 5$ GeV$^{-1}$ from the dilute gas approximation \cite{SVZ}, which is not favoured by the previous result.
\section{$\overline{m}_b(\overline{m}_b)$ from bottomium  }
\label{sec:mb}
\begin{figure}[hbt]
\begin{center}
\includegraphics[width=9cm]{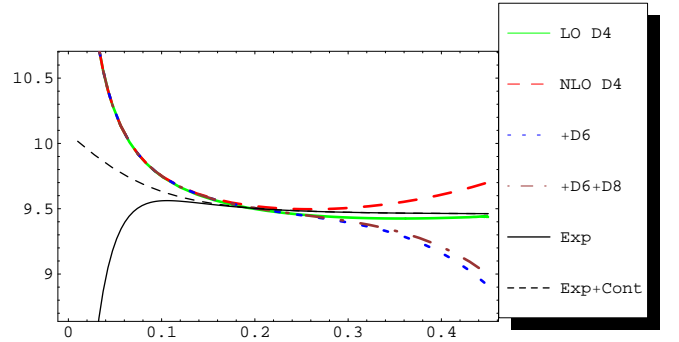}
\caption{\footnotesize Behaviour of the ratio of moments $\sqrt{{\cal R}^b_0}(\overline{m}_b^2)$ in GeV versus $\tau$ in GeV$^{-2}$ and for $\overline{m}_b(\overline{m}_b) = 4212$ MeV. The same legend as  in Fig. \ref {fig:ratioc} for the QCD contributions. The black continuous (rep. short dashed) curves are the experimental contribution including (resp. without) the QCD continuum.  } 
\label{fig:ratiob}
\end{center}
\end{figure}

\vspace*{-0.25cm}
 \nin
 Due to the  smaller values of $\alpha_s(m_b)$ and of $\tau$ at which the
 ratio of sum rules stabilizes  (see Fig. \ref{fig:ratiob}), the convergence of the QCD PT and NP series
 is faster here than for the case of charmonium. More explicitly, for $\tau=0.2$ GeV$^{-2}$, the PT expression of the ratio of sum rules reads numerically:
 \beq
\sqrt{{\cal R}_0}(\overline{m}_b^2)\vert_{PT}\simeq1-0.14a_s-0.35a_s^2-0.76a_s^3~,
\eeq
where we have used the QCD inputs in Eqs. (\ref{eq:cond_mom}),  (\ref{eq:mass_mom}) and (\ref{eq:rhoc}). We shall use the value of $\alpha_s$ deduced from the one from $\tau$-decay \cite{SNTAU,BNP}:
\beq
\alpha_s(m_b)=0.219(4).
\eeq
Using $\la \alpha_s G^2\ra$ from  Eq. (\ref{eq:cond_mom}), the improved  value of the triple gluon condensate in  Eq. (\ref{eq:rhoc}) from exponential charmonium sum rule and the modified factorized value of the $D=8$ dimension condensates, we reconsider here the determination
 of $\overline{m}_b(\overline{m}_b)$ from bottomium by adjusting the minimum (inflexion point) of  the QCD value of $\sqrt{{\cal R}^b_0}(\overline{m}_b^2)$ with the one from the data. This is achieved by the value of the running mass in GeV:
 \beq
\overline{m}_b(\overline{m}_b) = 4212(18)_\tau(0.2)_{\alpha_s}(25)_\nu(1.5)_{G^2+G^3+G^4}~(8)_{Coul},
\lb{eq:mba}
 \eeq
corresponding to $\tau=(0.20\pm 0.06)$ GeV$^{-2}$ where the inflexion point of the theoretical curve meets the experimental one  (see Fig. \ref{fig:ratiob}). \\
\b We notice that the sum of the NP condensate contributions is about 1.9 per mil of $m_b$ which
is quite small. \\
\b The error due to the $\alpha_s^{n\geq 4}$ is obtained by taking the $\pm$ sign in the coefficient of $\alpha_s^3$. The one due to $\nu$ is induced by moving  it in the range $(\overline{m}_b\pm 1)$ GeV. The one due to $G^4$ is the deviation from the modified to the usual factorization.  The one from Coulombic corrections has been estimated like in \cite{SNcb2} from the
  resummed expression of the spectral function \cite{EICHTEN} and adding to it the PT expression up to 
  order $\alpha_s^2$ computed in \cite{CHET2} which contains the familiar $(1-4C_Fa_s)$ factor due to the quarkonium annihilation through a single (transverse) virtual gluon and corrections to order v and logv for small v. Then, we deduce the final value:
    \beq
\overline{m}_b(\overline{m}_b) = 4212(32)~{\rm MeV}~,
\lb{eq:mb}
 \eeq
 which is slightly higher and less accurate  than the one from the $Q^2$-moments given in Eq.~(\ref{eq:mass_mom}) \cite{SNcb,SNcb2} but the two results are consistent within the errors.\\
\b By comparing the two results from the  $Q^2$-moments  and exponential sum rules, we are tempted to take the average of the results in Eqs. (\ref{eq:mass_mom}) and (\ref{eq:mb}) as a final result from the sum rules approaches:
  \beq
\overline{m}_b(\overline{m}_b) = 4177(11)~{\rm MeV}~.
\lb{eq:mbfinal}
 \eeq 
\nin
\section{Conclusions and outlook  }
\label{sec:conc}
\vspace*{-0.25cm}
 \nin
\b We have reconsidered the QCD-exponential sum rules by including higher order PT corrections to order $\alpha_s^3$ and non perturbative gluon condensate contributions up to $D=8$. We have also included the radiative corrections to the $D=4$ contribution. By studying the convergence of the PT series, we have concluded that the ratio of sum rules ${\cal R}_0$ in Eq. (\ref{eq:ratio}) is more convergent than the moment ${\cal L}_0$  in Eq. (\ref{eq:mom}).\\
\b From charmonium ratio of sum rules ${{\cal R}^c_0}(\overline{m}_c^2)$, we have deduced the ratio between the gluon condensates $\la \alpha_s G^2\ra$ and $\la g^3f_{abc} G^3\ra$ with the results in Eqs. (\ref{eq:rhoc}) and (\ref{eq:cond}), where we have used as inputs the values of  $\overline{m}_c(\overline{m}_c)$ from $Q^2$-moments in Eq. (\ref{eq:mass_mom}) \cite{SNcb} and the gluon condensate $\la \alpha_s G^2\ra$ from $Q^2$-moments \cite{SNcb} and from some other sources \cite{SNGh,SNI}. We have also used the modified factorization for the $D=8$ condensates proposed by \cite{BAGAN}, which was favoured in the analysis of $Q^2$-moments \cite{SNcb}. The present result improves earlier ones obtained in Eq. (\ref{eq:cond_mom}) from $Q^2$-moments \cite{SNcb}. \\
\b Using the previous results to the bottomium systems, we obtain, from ${{\cal R}^b_0}(\overline{m}_b^2)$, the running $b$-quark mass in Eq.~(\ref{eq:mb}) from the ratio of exponential sum rules. The result is slightly higher than the one from $Q^2$-moments in Eq. (\ref{eq:mass_mom}) and less accurate but agrees with it within the errors.
We are tempted to consider as a final result of $\overline{m}_b(\overline{m}_b)$ from the sum rules approach, the average given in Eq. (\ref{eq:mbfinal}) from the ratio of $Q^2$-moments  and that of exponential sum rules. 
The previous result improves older findings for the $b$-quark mass from bottomium systems by the author \cite{SNmc}, by Bell-Bertlmann in a series of papers \cite{BELL,BERT,NEUF} and by \cite{SHAW}. 


 \bibliography{mybib}
\end{document}